%% file: main.tex
\documentclass[10pt,runningheads]{llncs}
\usepackage{graphicx}
\usepackage{xr-hyper}
\usepackage{hyperref}
\usepackage{booktabs}
\usepackage{multirow}

\usepackage{amsmath}
\usepackage{amssymb}
\usepackage{amsfonts}
\usepackage{bm}

\usepackage{algorithm,algorithmic}
\renewcommand{\algorithmiccomment}[1]{\bgroup\hfill$\triangleright$~#1\egroup}

\usepackage[margin=1in]{geometry}  
\hypersetup{
    colorlinks=true,       
    linkcolor=blue,        
    citecolor=magenta,       
    filecolor=green,     
    urlcolor=cyan          
}

\usepackage[normalem]{ulem}
\usepackage{fancyhdr}
\newcommand{\E}{\mathbb{E}}

\newcommand{\KL}{\operatorname{KL}}

\newcommand{\CloneOT}{CloneOT}
\newcommand{\LINOT}{\CloneOT}   

\input{tables/numbers}

\input{tables/biology_numbers}


\begin{document}
\title{CloneOT: predicting clone-specific perturbation responses with lineage-supervised neural inverse optimal transport}
\titlerunning{CloneOT}
\author{Yanshuo Chen\inst{1, 2} \and Wei Chen\inst{3, 4} \and Heng Huang\inst{1, 2, 5}}
\authorrunning{Y. Chen et al.}
\institute{
Department of Computer Science, University of Maryland, College Park, MD, 20742, USA. \and
Center of Bioinformatics and Computational Biology, College Park, MD, 20740, USA. \and
Department of Biostatistics, University of Pittsburgh, Pittsburgh, PA, 15261, USA. \and
Department of Pediatrics, UPMC Children's Hospital of Pittsburgh, Pittsburgh, PA, 15224, USA. \and
Institute of Health Computing, University of Maryland, North Bethesda, MD, 20852, USA.}
\maketitle              
\begin{abstract}
\input{MainText/abstract}
\end{abstract}


\newpage
\setcounter{page}{1}

\section{Introduction}
\input{MainText/intro}

\section{Methods} \label{methods}
\input{MainText/methods}

\section{Results}
\input{MainText/results}

\section{Discussion}
\input{MainText/discussion}

\section{Acknowledgements}
We thank the authors of the Watermelon, ClonMapper and ReSisTrace studies for making their lineage-tracing data publicly available.

\section{Author contributions}
Y.C. conceived the method, implemented the software and performed the experiments. W.C. and H.H. supervised the project. All authors wrote and reviewed the manuscript.

\section{Data availability}
All datasets analyzed in this study are publicly available from the Gene Expression Omnibus under accession numbers GSE150949 (Watermelon \cite{watermelon}), GSE291678 (ClonMapper \cite{tnbc}) and GSE223003 (ReSisTrace \cite{resistrace}). The MSigDB gene sets are available at \url{https://www.gsea-msigdb.org}.

\section{Code availability}
The implementation of \LINOT, with its documentation, tutorials and the scripts of the simulated examples, is available at \url{https://github.com/poseidonchan/CloneOT}.

\bibliographystyle{splncs04}
\bibliography{mybib}
\end{document}

%% file: tables/numbers.tex
\newcommand{\trainminutes}{1.6}

%% file: tables/biology_numbers.tex
\newcommand{\bioClones}{1356}

\newcommand{\bioImproved}{eight}

\newcommand{\bioPrograms}{ten}

\newcommand{\bioScenarios}{12}

%% file: MainText/abstract.tex
Predicting single-cell perturbation responses requires mapping between two unpaired populations of cells measured before and after the perturbation. Optimal transport (OT) provides a principled framework for constructing such mappings by minimizing the transport cost, and neural OT methods (\textit{e.g.}, CellOT) have been employed for this prediction task. However, these methods rely on a fixed transport cost, typically the squared Euclidean distance. This cost may connect cells with the wrong responses. Lineage tracing provides the information needed to learn the cost, as heritable barcodes link the cells of the same clone before and after the perturbation. Here, we propose \CloneOT, a neural inverse OT method that learns both the transport cost and the transport map from partially lineage-traced single-cell datasets. \CloneOT\ represents the cost as a neural network. It learns the cost by minimizing the extra cost that the lineage adds to the optimal transport, so that the cheapest transport keeps the cells of each clone together. Our experiments demonstrate that \CloneOT\ recovers the correct transport on 2D datasets, while the squared Euclidean cost fails. Moreover, \CloneOT\ outperforms five baselines on 14 lineage-tracing perturbation scenarios of cancer cells, for both seen and unseen clones. Furthermore, we show that \CloneOT\ reveals the clone-specific transcriptional programs of drug-tolerant persister cells and forecasts the fate of each clone from its cells before treatment. Our implementation is available at \url{https://github.com/poseidonchan/CloneOT}.

%% file: MainText/intro.tex
Single-cell perturbation experiments measure how drugs, genetic interventions and other stimuli change the molecular states of individual cells. They have important implications for cancer research \cite{cancer} and drug development \cite{drug}. Predicting these responses computationally is also a key task of recent efforts toward virtual cell models \cite{bunne2024virtual,roohani2025virtual,adduri2025state}. In particular, predicting the response of a cell from its state before the perturbation can explain why cells of the same population respond differently. For example, a small fraction of cancer cells persists under targeted therapy \cite{watermelon,fatemap}. However, single-cell RNA sequencing (scRNA-seq) destroys the cells it measures. It is therefore impossible to observe the same cell before and after the perturbation, and a model must learn a mapping between two unpaired populations.

Since no cell is observed twice, a model has to decide which cell before the perturbation becomes which cell after it. Previous models make this decision implicitly, by making different assumptions during modeling. scGen assumes that the perturbation shifts every cell by the same vector in a learned latent space \cite{scgen}, and SquidDiff, a conditional diffusion model, applies the same shift to its conditioning vector \cite{squidiff}. chemCPA and bioLORD assume that the basal state of a cell and the effect of the perturbation can be disentangled, so the perturbation changes the effect while the cell keeps its basal state \cite{chemcpa,biolord}. Optimal transport (OT) methods such as CellOT assume that all features of a cell are equally important under the perturbation and they move cells onto the target population at the minimal total cost measured by the squared Euclidean distance \cite{cellot,w1ot,cmonge}. This assumption does not match the observation that a perturbation changes some features much more than others. However, unpaired data produced by conventional scRNA-seq cannot \emph{identify} the correct cost. For example, if two equally abundant groups of cells exchanged their responses, the treated population would look exactly the same. Hence, an optimal transport model with the Euclidean cost would then assign both groups the wrong response.

Lineage tracing provides the missing link. Heritable barcodes mark the descendants of a common ancestor. The cells of a clone can therefore be identified before and after treatment, even though their states and numbers change \cite{watermelon,tnbc,resistrace,fatemap}. A clone links a set of source cells to a set of target cells without pairing individual cells. 
Previous methods used lineage information to constrain the couplings of OT-based trajectory inference \cite{wot,moscot} and to supervise transition maps \cite{lineageot,moslin,cospar,superot}. Nevertheless, these methods relate the cells under a fixed cost or transition model. They do not learn the cost itself.

Learning a cost from observed transport is the subject of inverse OT \cite{stuart2020inverse,galichon2022cupid,li2019learning}. When matched pairs are observed, the cost can be estimated by maximizing their likelihood under the entropic coupling \cite{dupuy2014personality}. OT-SI is the closest method to lineage tracing, as it learns a cost from correspondences between subsets of two point clouds \cite{otsi}. However, these methods are discrete OT solvers, so they do not learn a map that predicts the responses of new cells.

To address these limitations, we present \CloneOT, a neural inverse OT method that learns both the transport cost and the transport map from partially lineage-traced single-cell datasets (Fig.~\ref{fig:overview}). Unlike previous methods, \CloneOT\ uses the available lineage information to learn the transport cost and the corresponding neural optimal transport map to predict the response of a new cell from its expression alone. \CloneOT\ therefore uses lineage tracing to learn how the response differs between clones.

We demonstrate that \CloneOT\ is capable of 1) recovering the correct transport on two 2D datasets, while the squared Euclidean cost fails; 2) predicting the responses of lineage-traced cancer cells in 14 scenarios more accurately than five baselines, for both seen and unseen clones; and 3) revealing the clone-specific programs of drug-tolerant persister cells, such as the interferon and antioxidant programs of non-cycling and cycling persisters, and forecasting the fate of each clone from its cells before treatment.

%% file: MainText/methods.tex
\subsection{Overview}
\label{sec:overview}

\begin{figure}[!t]
\centering
\includegraphics[width=\textwidth]{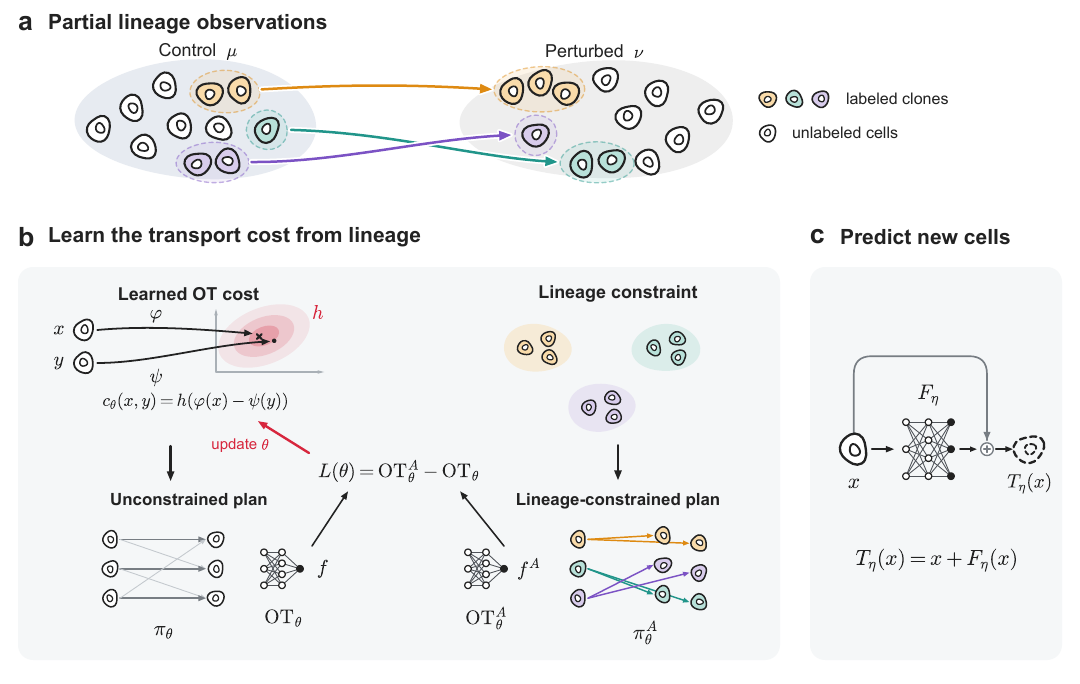}
\caption{\textbf{Overview of \CloneOT.} \textbf{a.} Lineage barcodes identify, for a fraction of the clones (colored), which source and target cells belong to the same clone. \textbf{b.} The transport cost $c_\theta(x,y)=h_\theta(\varphi_\theta(x)-\psi_\theta(y))$ is a neural network with neural ODE flows $\varphi_\theta,\psi_\theta$ and a symmetric input convex network $h_\theta$. Training alternates between solving the entropic OT problems without and with the lineage constraint by neural dual potentials and descending $L(\theta)=\mathrm{OT}^A_\theta-\mathrm{OT}_\theta$, the extra cost of respecting the lineage. \textbf{c.} A neural map $T_\eta$ amortizes the learned coupling and predicts the response of a new cell.} \label{fig:overview}
\end{figure}

Single-cell perturbation experiments measure one population of cells before a perturbation, the source cells, and another population after it, the target cells. Sequencing destroys the cells it measures, so we never observe how an individual source cell responds. The task is to learn a map that predicts the response of any source cell from its expression (Fig.~\ref{fig:overview}a). Optimal transport (OT) addresses this task by moving the source population onto the target population at the smallest total cost. The transport cost $c(x,y)$ specifies how expensive it is for a cell in state $x$ to become a cell in state $y$. The cost therefore decides which source cells are connected with which responses. Most OT methods fix the cost to the squared Euclidean distance. This choice assumes that cells change as little as possible in every direction of the expression space. The assumption fails when a perturbation changes some directions much more than others, and unpaired data alone cannot reveal the correct cost.

Lineage tracing provides the information needed to learn the cost. Heritable barcodes mark the descendants of each ancestor cell. For a fraction of the clones, we therefore know which source cells and which target cells belong together, although we do not know which source cell gave rise to which target cell (Fig.~\ref{fig:overview}a). \CloneOT\ uses this information with a simple idea. Under a good cost, the cheapest transport of the source population onto the target population should already keep the cells of each clone together. \CloneOT\ therefore solves two transport problems with the same cost, one without and one with the lineage as a constraint, and measures the extra cost that the lineage adds (Fig.~\ref{fig:overview}b). If the lineage-consistent transport is much more expensive than the best transport, the cost disagrees with the lineage, and \CloneOT\ changes the cost to reduce this gap. The cost is a neural network, so the cost of a change can depend on the state of a cell. Both transport problems are also solved with neural networks, so the cost is learned without differentiating through an iterative solver. After training, a neural map learns the expected response of each cell under the learned transport. The map predicts the response of a new source cell from its expression alone, without its barcode (Fig.~\ref{fig:overview}c).

\subsection{\CloneOT}
\label{sec:linot}

\subsubsection{Problem formulation}
The training data consist of source cells $\{x_i\}_{i=1}^N$ and target cells $\{y_j\}_{j=1}^M$, with empirical distributions $\mu$ and $\nu$. Each cell is represented by the posterior mean of a variational autoencoder (VAE) \cite{vae}. For each labeled clone $k=1,\dots,K$, we know its source cells $S_k$ and its target cells $Y_k$. All other cells, including the clones observed on only one side, form the unlabeled sets $S_0$ and $Y_0$. The goal is a map that predicts the response of a new source cell from its expression alone.

\subsubsection{Neural transport cost}
\CloneOT\ parameterizes the cost as a neural network of the form proposed by Howard et al.\ \cite{howard2024differentiable},
\begin{align}
c_\theta(x,y)=h_\theta\big(\varphi_\theta(x)-\psi_\theta(y)\big),\qquad h_\theta(d)=g_\theta(d)+g_\theta(-d)+\frac{\alpha}{2}\|d\|^2,
\label{eq:cost}
\end{align}
in coordinates centered at the source and target means. The maps $\varphi_\theta$ and $\psi_\theta$ are neural ODE flows \cite{chen2018neural,grathwohl2019ffjord}. Specifically, $\varphi_\theta(x)$ is the state at time 1 of $\dot z=v_\theta(z,t)$ started at $z(0)=x$, where $v_\theta$ is a neural velocity field, and $\psi_\theta$ has its own velocity field. The function $g_\theta$ is an input convex neural network (ICNN) \cite{amos2017input} with softplus activations and nonnegative weights between its hidden layers. Hence $h_\theta$ is convex, symmetric and minimal at $d=0$, and the quadratic term with $\alpha>0$ makes it strictly convex. The flows are invertible, so the cost compares the two populations in learned coordinates. A target state $y$ is cheap for a source cell $x$ when $\psi_\theta(y)$ is close to $\varphi_\theta(x)$ in the geometry of $h_\theta$. Moreover, the cost satisfies the twist condition, because the flows are diffeomorphisms and $h_\theta$ is strictly convex. For a source distribution with a density, the unregularized optimal transport is therefore a unique map of the form $\psi_\theta^{-1}\circ T\circ\varphi_\theta$, and $T$ is the optimal map for the cost $h_\theta$ between the transformed populations \cite{gangbo1996geometry,howard2024differentiable}. With $\varphi_\theta=\psi_\theta=\mathrm{id}$ and $h_\theta(d)=\frac12\|d\|^2$, $c_\theta$ is the squared Euclidean cost of $W_2$ OT. \CloneOT\ starts near this cost, with zero velocity fields and small output weights of the ICNN, and moves away from it as the lineage requires.

\subsubsection{Learning the cost from lineage}
Clones expand and shrink during an experiment. \CloneOT\ therefore gives the source cells of labeled clone $k$ the total mass $\nu(Y_k)$, the share of the clone among the target cells. The unlabeled source cells share the remaining mass equally, so the source distribution is
\begin{align}
\omega=\sum_{k=1}^K\nu(Y_k)\,\mu(\cdot\mid S_k)+\nu(Y_0)\,\mu(\cdot\mid S_0).
\label{eq:weights}
\end{align}
The coupling of the two populations under $c_\theta$ is the solution $\pi_\theta$ of the entropic OT problem
\begin{align}
\mathrm{OT}_\theta=\min_{\pi\in\Pi(\omega,\nu)}\ \int c_\theta\,d\pi+\varepsilon\KL(\pi\,\Vert\,\omega\otimes\nu).
\label{eq:eot}
\end{align}
Here $\Pi(\omega,\nu)$ is the set of couplings with marginals $\omega$ and $\nu$, $\KL$ is the Kullback--Leibler divergence and $\varepsilon>0$ is a temperature \cite{cuturi2013sinkhorn,computationalOT}. Under a good cost, this coupling should already connect the source cells of each labeled clone with the target cells of the same clone. \CloneOT\ therefore compares it with the same problem under the lineage constraint. Under this constraint, a target cell of labeled clone $k$ may only descend from source cells of clone $k$, and an unlabeled target cell may only descend from unlabeled source cells. We write $\mathrm{OT}^A_\theta$ and $\pi^A_\theta$ for the optimal value and coupling of the constrained problem. They exist because Eq.~(\ref{eq:weights}) gives every clone exactly its share of the target cells. A constraint can only increase the optimal cost, so the difference
\begin{align}
L(\theta)=\mathrm{OT}^A_\theta-\mathrm{OT}_\theta\ \ge\ 0
\label{eq:likelihood}
\end{align}
is never negative. It is close to zero when the unconstrained coupling nearly respects the lineage, and it grows with the extra cost that the lineage adds to the transport. \CloneOT\ learns the cost by minimizing $L/\varepsilon$. Equivalently, $L/\varepsilon$ is the smallest Kullback--Leibler divergence between a lineage-consistent coupling and $\pi_\theta$. Minimizing it therefore moves the coupling that the cost predicts toward the observed lineage. When every clone consists of one source cell and one target cell, $L/\varepsilon$ is the negative log-likelihood of the observed pairs, the classical inverse OT estimator \cite{dupuy2014personality}. Supplementary Section~\ref{S-sec:theory} proves these properties. The derivative of an optimal transport cost with respect to a parameter of the cost is the expected derivative of the cost under the optimal coupling \cite{danskin1966theory}. Hence
\begin{align}
\nabla_\theta L(\theta)=\E_{\pi^A_\theta}\big[\nabla_\theta c_\theta\big]-\E_{\pi_\theta}\big[\nabla_\theta c_\theta\big],
\label{eq:gradient}
\end{align}
and the gradient only requires expectations under the two couplings.

\subsubsection{Neural entropic optimal transport solver}
Neural OT solvers represent the dual potentials of a transport problem by neural networks \cite{makkuva2020optimal,amos2023amortizing}, and CellOT uses them to predict perturbation responses with a fixed cost \cite{cellot,w1ot}. \CloneOT\ solves its two entropic problems in the same way. Supplementary Section~\ref{S-sec:background} reviews inverse OT methods. For a potential $f$ on the source cells, the best potential on the target cells is the entropic $c$-transform
\begin{align}
g_f(y)=-\varepsilon\log\int\exp\Big(\frac{f(x)-c_\theta(x,y)}\varepsilon\Big)\,d\omega(x).
\label{eq:ctransform}
\end{align}
The $c$-transform has a closed form, so only $f$ has to be learned, and
\begin{align}
\mathrm{OT}_\theta=\max_f D_\theta(f),\qquad D_\theta(f)=\int f\,d\omega+\int g_f\,d\nu .
\label{eq:dual}
\end{align}
The constrained problem has the same form, $\mathrm{OT}^A_\theta=\max_f D^A_\theta(f)$, but the integral in its $c$-transform runs only over the source cells allowed for $y$. At the optimum, a target cell $y$ descends from the source cell $x$ with probability proportional to $e^{(f(x)-c_\theta(x,y))/\varepsilon}\,\omega(x)$ among the allowed source cells. ICNN-based neural OT solvers play a min-max game between two networks \cite{makkuva2020optimal}. In contrast, each problem of \CloneOT\ needs a single network, and its $c$-transform needs no inner optimization. \CloneOT\ parameterizes the two potentials as $f(x)=s_\theta f_\xi(\varphi_\theta(x))$ and $f^A(x)=s_\theta f^A_{\xi'}(\varphi_\theta(x),e_{k(x)})$. Here $s_\theta$ is the median of $c_\theta$ over random pairs of training cells, so the networks work in units of the cost, and $e_{k(x)}$ is a learned embedding of the labeled clone of $x$. Training alternates stochastic gradient steps (Supplementary Algorithm~\ref{S-alg:linot}). First, the potentials ascend $D_\theta$ and $D^A_\theta$ on minibatches of 4,096 target cells, and the $c$-transform of each target cell is computed exactly over all training source cells. Then the cost descends $L/\varepsilon$ with the potentials held fixed. \CloneOT\ estimates the two expectations of Eq.~(\ref{eq:gradient}) with source cells sampled from the conditional distributions of the two couplings. In each cost update, it draws 1,024 target cells and samples eight source cells for each of them from each coupling.

\subsubsection{Neural map and prediction}
Once the cost is learned, the predicted response of a source cell is the average of its possible descendants under $\pi_\theta$,
\begin{align}
\E_{\pi_\theta}[y\mid x]=\sum_{j=1}^M\frac{\exp\{(g_f(y_j)-c_\theta(x,y_j))/\varepsilon\}}{\sum_{l}\exp\{(g_f(y_l)-c_\theta(x,y_l))/\varepsilon\}}\,y_j .
\label{eq:barycentric}
\end{align}
This prediction is defined for any $x$, but it requires all training target cells. \CloneOT\ therefore amortizes the prediction with a neural map $T_\eta(x)=x+F_\eta(x)$, fitted by least squares on minibatches of training source cells \cite{seguy2018large}. On single-cell data, $F_\eta$ is fitted so that the decoded map $D(T_\eta(x))$ matches the expected decoded descendant $\E_{\pi_\theta}[D(y)\mid x]$, where $D$ is the VAE decoder. \CloneOT\ then predicts the expression profile $D(T_\eta(x))$ of a new source cell without its clone or any target measurement.

\subsection{Implementation and baselines}
\label{sec:baselines}
\label{sec:training}
The potentials and the map are multilayer perceptrons with three hidden layers of width 128. The velocity fields of the flows and the ICNN have two hidden layers of width 64. All networks are trained with Adam. The temperature is $\varepsilon=0.1\,m$, where $m$ is the median of $\frac12\|y-x\|^2$ over random pairs of training cells. The same settings are used for the 2D datasets and all single-cell tasks (Supplementary Table~\ref{S-tab:settings}). On single-cell data, the map checkpoint is selected on the validation clones. The selection criterion is the mean of $\ell_2$, $1-r^2$ and the energy score over the validation clones, each relative to the prediction of every clone by the mean of the training target cells. All models were trained on one NVIDIA RTX 5090 GPU, and \CloneOT\ trains in a median of \trainminutes\ minutes per model (Supplementary Table~\ref{S-tab:runtime}).

We compare \CloneOT\ with five published methods. They are the neural OT method CellOT \cite{cellot}, the diffusion model SquidDiff \cite{squidiff} and three generative perturbation models, scGen \cite{scgen}, chemCPA \cite{chemcpa} and bioLORD \cite{biolord}. Each method is trained per task with its recommended settings on the same cells and returns one predicted expression profile per source cell. Deterministic methods return their output, chemCPA and bioLORD return the mean of their predicted distribution, and SquidDiff returns one generated profile. Each method keeps the checkpoint with the best energy score on the validation clones, and none of them uses lineage information in training. As in CellOT \cite{cellot,w1ot}, we also report two naive baselines. \emph{Identity} predicts every source cell by its own expression, and \emph{Observed} predicts every clone by the training target cells. Observed is not an upper bound in the clone-level evaluation, since its clone means are those of the whole target population. On the 2D datasets, we also compare with OT-SI \cite{otsi}, the inverse OT method closest to \CloneOT, and with neural W2OT, the solver underlying CellOT \cite{cellot,amos2023amortizing}. OT-SI is a discrete OT solver without a map for new points, so we extend its plan to new points through its dual potential and predict them by the barycentric projection. Supplementary Section~\ref{S-sec:baseline-settings} lists the settings of all baselines.

\subsection{Datasets and evaluation}
\label{sec:datasets}
\label{sec:metrics}
The 2D datasets are the T-shape dataset and the rotated-moons dataset. Methods are evaluated on 400 new source points by the fraction of predictions on the correct class and by the mean distance to the true target position (Supplementary Section~\ref{S-sec:datasets}). The single-cell data comprise 14 scenarios from three studies. Twelve time and sorting scenarios come from Watermelon lineage tracing of PC9 lung cancer cells treated with osimertinib \cite{watermelon}. One scenario comes from doxorubicin treatment of MDA-MB-231 breast cancer cells with ClonMapper barcodes \cite{tnbc}, and one from olaparib treatment of Kuramochi ovarian cancer cells in ReSisTrace \cite{resistrace}. Expression is normalized and log-transformed, and 2,000 highly variable genes are selected on the training cells \cite{scanpy,seuratv3}. A VAE with the architecture of W1OT \cite{w1ot}, hidden layers of widths 256 and 128 and a 50-dimensional latent space is then trained on the source and target cells of the training set. In the setting of \emph{seen clones} (IID), about 20\% of the cells of every training clone with at least two cells on each side are held out. These evaluation clones are split at random into a validation half for model selection and a test half for evaluation. In the setting of \emph{unseen clones} (OOD), entire clones are held out, so the test clones never appear in training. In each setting, 10\%, 20\%, 40\% or 80\% of the training clones are labeled, in nested sets.

Following CellOT \cite{cellot,w1ot}, we evaluate predictions on the top 50 differentially expressed genes of the training cells, separately for every evaluation clone. The metrics are (i) the $\ell_2$ distance between the predicted and observed clone-mean expression, (ii) the squared Pearson correlation $r^2$ between the two clone means, and (iii) the energy score \cite{gneiting2007strictly}. For a clone with predicted profiles $\{p_i\}_{i=1}^n$ and observed target cells $\{y_j\}_{j=1}^m$, the energy score is
\begin{align}
\mathrm{ES}=\frac1{nm}\sum_{i,j}\|p_i-y_j\|-\frac1{2n^2}\sum_{i,i'}\|p_i-p_{i'}\|.
\label{eq:energy}
\end{align}
It is a strictly proper scoring rule, so its expectation is lowest when the predicted profiles follow the true distribution of the cells of the clone. Scores are averaged over clones. Two methods are compared with the two-sided Wilcoxon signed-rank test on per-clone differences within each scenario, with Benjamini--Hochberg adjustment (Supplementary Section~\ref{S-sec:evaluation}).

%% file: MainText/results.tex
\subsection{\CloneOT\ learns the transport cost on 2D datasets}

\begin{figure}[!t]
\centering
\includegraphics[width=\textwidth]{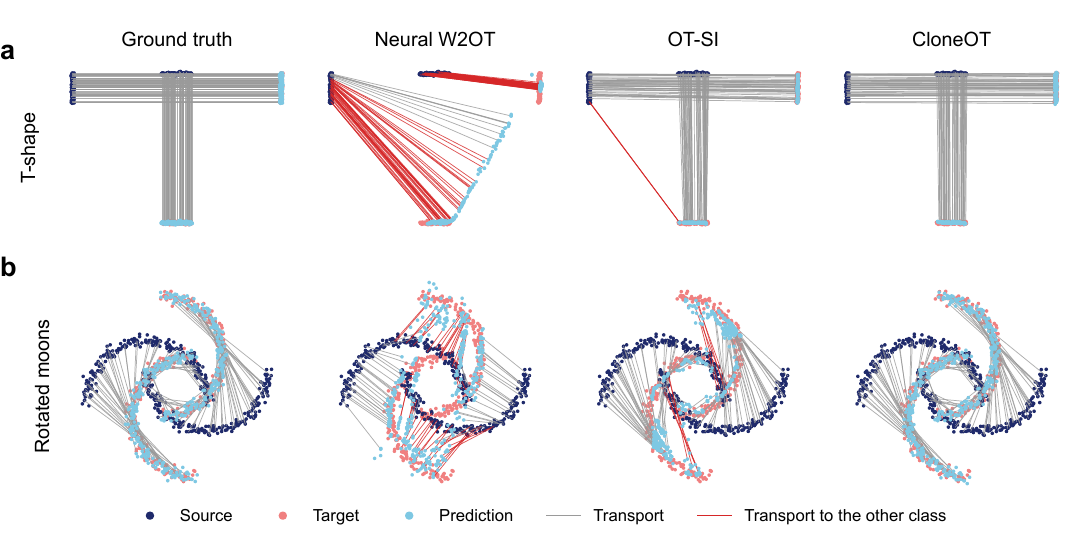}
\caption{\textbf{\CloneOT\ on 2D datasets.} Source points (dark blue), target points (pink) and predictions for new source points (light blue) are shown with transport rays for a quarter of the points. Red rays end on the wrong class. \textbf{a.} The T-shape dataset \cite{howard2024differentiable}. For 20\% of the source points, the corresponding target points are given. \textbf{b.} The rotated-moons dataset \cite{otsi}. The only correspondence is between each source moon and its rotated moon. Neural W2OT uses the squared Euclidean cost, and all methods are shown for training seed 0.} \label{fig:examples}
\end{figure}

We first validate our model on 2D datasets (Section~\ref{sec:datasets} and Fig.~\ref{fig:examples}). We use these datasets to examine whether the transport cost needs to be learned, and whether \CloneOT\ learns it more accurately than a discrete inverse OT solver. In both datasets, the supervision is a set of correspondences between subsets of source and target points, and these subsets play the role of clones. We compare \CloneOT\ with two methods. Neural W2OT uses the squared Euclidean cost and is the solver underlying CellOT \cite{cellot,amos2023amortizing}, and OT-SI is a discrete inverse OT solver \cite{otsi}.

First, we test the T-shape dataset \cite{howard2024differentiable}. In this dataset, a vertical bar on the left moves to the right, and a horizontal bar on top moves down (Fig.~\ref{fig:examples}a). Under the squared Euclidean cost, moving the left bar down and the top bar to the right is cheaper. Neural W2OT therefore sends almost all new points to the wrong bar. In contrast, \CloneOT\ learns the cost from the correspondences of 20\% of the source points, and it places every new point on the correct bar. This indicates that the cost must be learned when the squared Euclidean cost prefers a wrong transport.

Next, we compare \CloneOT\ with OT-SI on the rotated-moons dataset of OT-SI \cite{otsi}. The target is the two moons rotated by $90^\circ$. The only supervision is which source moon corresponds to which rotated moon (Fig.~\ref{fig:examples}b). Neural W2OT sends about half of the new points to the wrong moon. OT-SI places most points on the correct moon, but its accuracy changes with the training seed, and on the T-shape it fails for some seeds. The results show that \CloneOT\ places every new point on the correct moon for all seeds and recovers the rotation itself (Supplementary Table~\ref{S-tab:examples}). This indicates that \CloneOT\ learns the transport more accurately and more stably than the discrete inverse OT solver.

\subsection{\CloneOT\ accurately predicts clone-level responses}
\label{sec:benchmark}

\begin{figure}[!t]
\centering
\includegraphics[width=\textwidth]{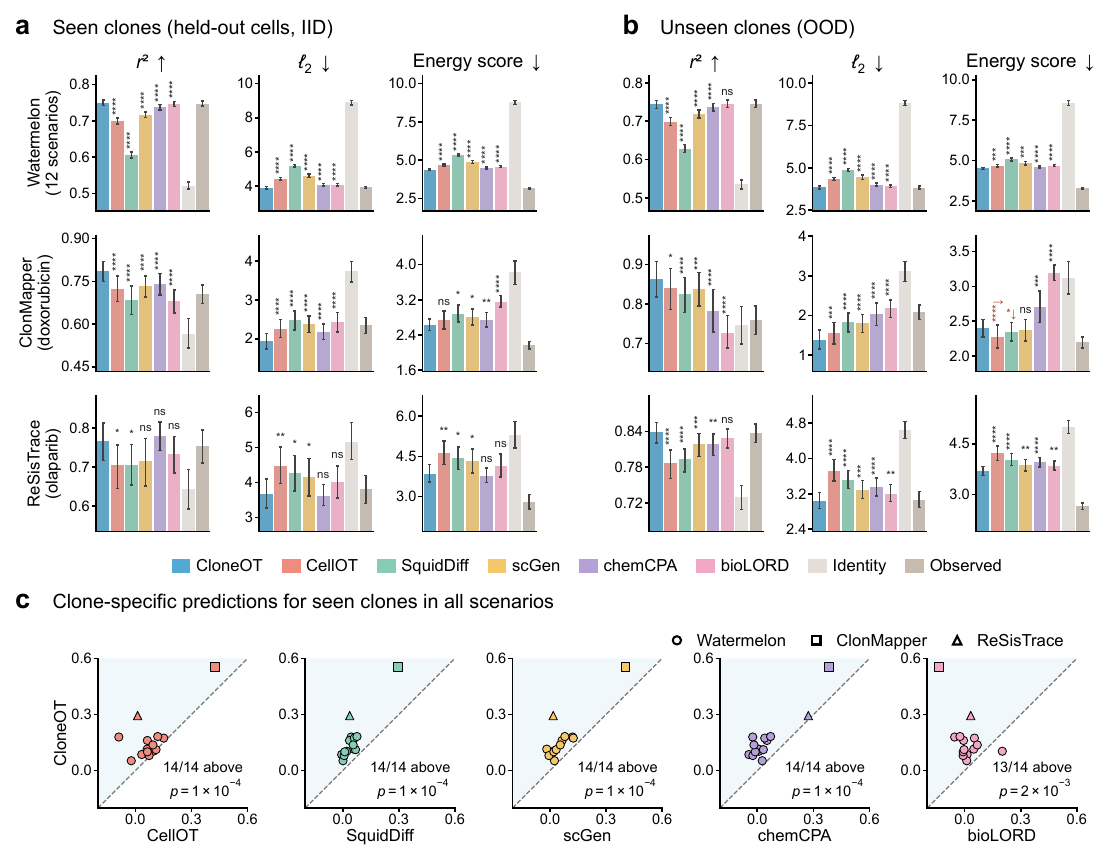}
\caption{\textbf{Performance benchmark with 80\% labeled clones.} \textbf{a.} Seen clones (IID setting). \textbf{b.} Unseen clones (OOD setting). Bars show means over the test clones of each study with 95\% bootstrap confidence intervals. Stars denote two-sided Wilcoxon signed-rank tests against \CloneOT\ over clones (*$p<0.05$, **$p<0.01$, ***$p<0.001$, ****$p<10^{-4}$, ns not significant), and red arrows mark methods significantly better than \CloneOT. \textbf{c.} Clone-specific correlation for seen clones, \CloneOT\ against each baseline with one point per scenario and $p$ from a paired Wilcoxon signed-rank test over scenarios.} \label{fig:benchmark}
\end{figure}

We apply \CloneOT\ to the 14 lineage-tracing scenarios of Watermelon \cite{watermelon}, ClonMapper \cite{tnbc} and ReSisTrace \cite{resistrace}. We compare it with the five baselines and the two naive baselines, Identity and Observed, using the per-clone $r^2$, $\ell_2$ and energy score (Sections~\ref{sec:baselines} and \ref{sec:metrics}). We consider two settings. For seen clones (IID setting), about 20\% of the cells of every training clone are held out, and each model predicts these cells from their expression before the perturbation. For unseen clones (OOD setting), entire clones are held out, so every evaluation clone is absent from training. In both settings, 10\%, 20\%, 40\% or 80\% of the training clones carry lineage labels.

The results show that \CloneOT\ is the most accurate of the six methods on all three metrics for seen clones, except in ReSisTrace, where it performs on par with chemCPA (Fig.~\ref{fig:benchmark}a). In paired tests over the clones of each scenario, \CloneOT\ is significantly more accurate than the baselines in most comparisons, and it is never significantly less accurate (Supplementary Table~\ref{S-tab:iid-summary}). The naive baselines bracket these results. Identity assumes no response and is the least accurate prediction. Observed predicts every clone by the measured target population, so its clone means are those of the average response. Observed also has the lowest energy score, because it consists of measured cells and therefore carries their cell-to-cell variability. Notably, in Watermelon, \CloneOT\ is the only method that predicts clone means more accurately than Observed. This can be attributed to the fact that the other methods mainly predict the response that all cells share, whereas \CloneOT\ also predicts how each clone deviates from it.

To examine the predictions themselves, we ask how well each method predicts the differences between clones (Fig.~\ref{fig:benchmark}c). For each test clone, we compute its clone-specific deviation. The deviation is the mean expression of the clone over the 50 evaluation genes minus the average of these clone means over all test clones of the scenario. We compute it once for the prediction and once for the observed target cells. The deviation removes the response that all treated cells share. Hence a method that predicts the same profile for every clone, such as Observed, has no deviations. We then define the clone-specific correlation $r$ as the Pearson correlation between the predicted and observed deviations over all clones and genes of a scenario. \CloneOT\ achieves a higher clone-specific correlation than every baseline in nearly all scenarios. This shows that the gains of \CloneOT\ extend beyond the shared response to the differences between clones.

We further evaluate \CloneOT\ on unseen clones (Fig.~\ref{fig:benchmark}b). \CloneOT\ again predicts the clone means with the lowest $\ell_2$ of the six methods, and it has the lowest energy score in Watermelon and ReSisTrace. In paired tests, it is rarely significantly less accurate than a baseline (Supplementary Table~\ref{S-tab:ood-summary}). These results indicate that the cost learned from the labeled clones generalizes to clones absent from training.

\subsection{\CloneOT\ improves with lineage supervision and transfers it to unlabeled clones}
\label{sec:supervision}

\begin{figure}[!t]
\centering
\includegraphics[width=\textwidth]{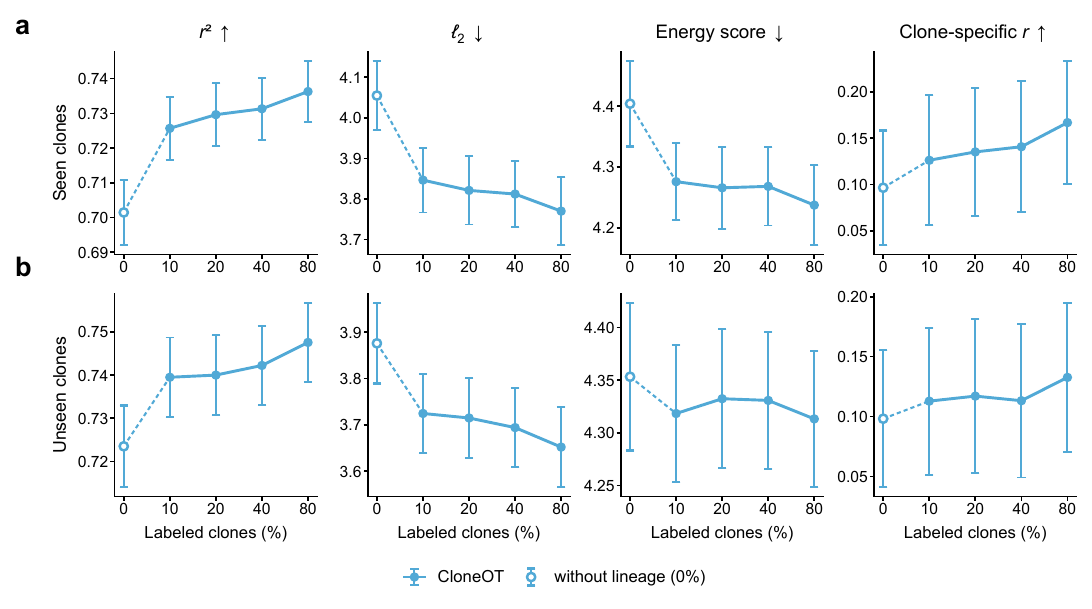}
\caption{\textbf{Lineage supervision improves \CloneOT.} \textbf{a.} $r^2$, $\ell_2$, energy score and clone-specific correlation (Fig.~\ref{fig:benchmark}c) for seen clones against the fraction of labeled training clones. The value at 0\% is \CloneOT\ without lineage (open circles). \textbf{b.} The same for unseen clones. Points are means over the 14 scenarios with 95\% confidence intervals over clones (over scenarios for the clone-specific correlation).} \label{fig:supervision}
\end{figure}

Next, we examine how the amount of lineage information affects \CloneOT. The results show that lineage improves every metric for both seen and unseen clones, and that the accuracy generally increases with the fraction of labeled clones (Fig.~\ref{fig:supervision}). Without lineage, \CloneOT\ is entropic $W_2$ OT. Like the other OT methods, it then predicts clone means less accurately than Observed. Learning the cost from only 10\% of the clones already improves every metric. This indicates that a few labeled clones are enough to improve the cost, so \CloneOT\ suits experiments in which only some clones are identified on both sides. With 80\% labeled clones, \CloneOT\ reduces $\ell_2$ in every scenario for both seen and unseen clones, and it has a significantly lower $\ell_2$ than the same solver with the fixed squared Euclidean cost in almost every scenario (Supplementary Table~\ref{S-tab:supervision-tests}). Removing the flows or replacing the ICNN by a quadratic function also reduces the accuracy (Supplementary Section~\ref{S-sec:ablations}). This indicates that a flexible neural transport cost is necessary.

Interestingly, the gains extend to the clones without labels. At every fraction of labeled clones, $\ell_2$ decreases at least as much on unlabeled as on labeled test clones. This can be attributed to the fact that all cells share the learned cost, so the information of the labeled clones transfers to the other clones.

\subsection{\CloneOT\ reveals the clone-specific programs and fates of drug-tolerant persister cells}
\label{sec:biology}

\begin{figure}[!t]
\centering
\includegraphics[width=\textwidth]{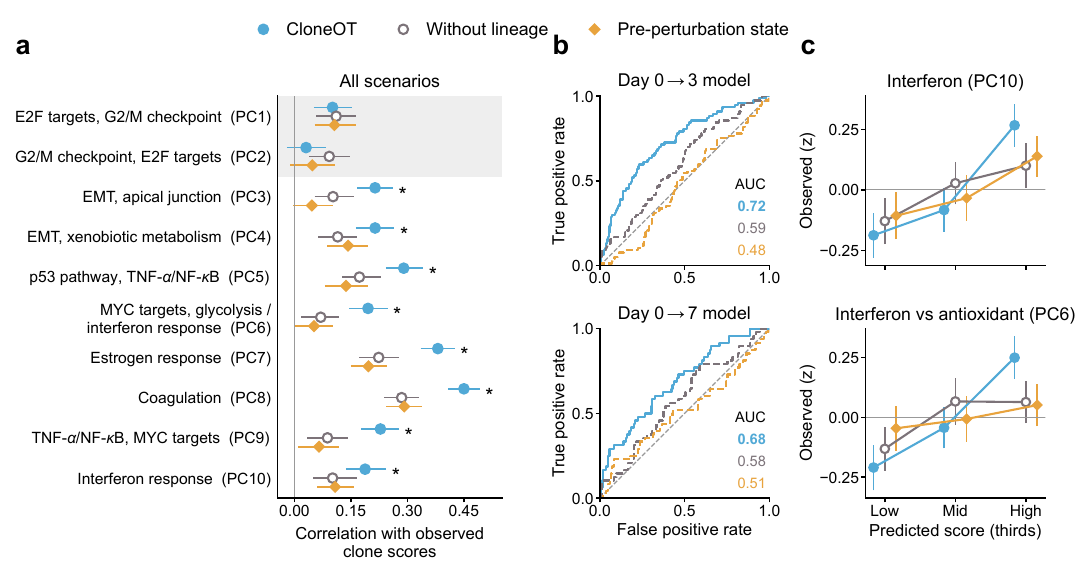}
\caption{\textbf{\CloneOT\ predicts the clone-specific programs and fates of drug-tolerant persister cells} (Watermelon, 80\% labeled clones). \textbf{a.} Correlation across seen test clones between the predicted and the observed program score, pooled over the \bioScenarios\ scenarios (\bioClones\ clones summed over scenarios) with 95\% bootstrap confidence intervals. Programs are the leading principal components of the post-treatment cells and are labeled by the Hallmark sets enriched at their two ends. Asterisks mark a gain from lineage (paired bootstrap, Benjamini--Hochberg $q<0.05$), and gray marks no gain. \textbf{b.} Forecast of the day-14 fate (cycling or non-cycling persisters) of day-0 clones from nine gene set scores by cross-validated logistic regression, with AUC averaged over 20 repetitions. \textbf{c.} Observed program score of the seen test clones in each third of the predicted score, with 95\% confidence intervals.} \label{fig:programs}
\end{figure}

\CloneOT\ predicts the differences between clones. We therefore ask what these differences mean biologically. In the Watermelon study, most clones gave rise to only cycling or only non-cycling persister cells \cite{watermelon}. Cycling persisters had higher NRF2 and glutathione antioxidant signatures. Non-cycling persisters had higher interferon-$\alpha$, cholesterol homeostasis and Notch signatures. None of these programs was upregulated before treatment. We therefore ask which programs \CloneOT\ predicts better with lineage, and whether \CloneOT\ forecasts the fate of a clone and its programs from the cells before treatment.

To answer these questions, we decompose the post-treatment cells of the \bioScenarios\ Watermelon scenarios into \bioPrograms\ transcriptional programs. The programs are the leading principal components of the training target cells, pooled over the scenarios. We annotate each program by the MSigDB Hallmark gene sets \cite{liberzon2015molecular,subramanian2005gene} enriched among its most positive and most negative genes (Supplementary Table~\ref{S-tab:programs}). For every test clone of the seen-clone setting, we score each program on its observed target cells, on the predictions for its source cells and on its source cells before the perturbation. We then measure how well each model predicts the program scores by the Pearson correlation across test clones, after standardizing the scores within each scenario (Fig.~\ref{fig:programs}a). The confidence intervals and the comparison with \CloneOT\ without lineage come from a paired bootstrap over clones (Supplementary Section~\ref{S-sec:program-tables}).

Lineage does not improve the prediction of the two leading programs, the E2F and G2/M cell-cycle programs. This is consistent with the cell-cycle phase being a transient state rather than a clonal one. In contrast, lineage significantly improves the prediction of the other \bioImproved\ programs, in most scenarios (Fig.~\ref{fig:programs}a and Supplementary Fig.~\ref{S-fig:program-gains}). Consistent with the original study, these programs include the ones that separate the persister fates. They include an interferon-response program (PC10) and a program that opposes interferon-response genes to MYC targets, glycolysis and antioxidant genes (PC6). They also include an EMT program with the NRF2 targets \textit{AKR1C2}, \textit{AKR1C3} and \textit{GSTM4} (PC4) and a p53 stress program (PC5). Scoring every MSigDB gene set separately gives the same pattern (Supplementary Table~\ref{S-tab:genesets}).

We next forecast the fate of each clone, the central question of the original study. A day-0 clone is cycling if its day-14 descendants appear only in the mCherry-low persister fraction, and non-cycling if they appear only in the mCherry-high fraction \cite{watermelon}. The mCherry-low fraction contains the cells that kept dividing. We describe each seen and unseen evaluation clone by nine fate-related MSigDB gene sets scored on its predicted response, including the NRF2, glutathione, interferon-$\alpha$ and cholesterol signatures. We then classify the two fates by logistic regression with five-fold cross-validation, repeated 20 times (Supplementary Section~\ref{S-sec:program-tables}). The results show that \CloneOT\ forecasts the fate clearly better than without lineage, from both its day-3 and its day-7 predictions (Fig.~\ref{fig:programs}b). In contrast, the same gene sets scored on the cells before treatment perform at the level of chance. This agrees with the original finding that the two fates arise from distinct lineages, and that their programs are not upregulated before treatment \cite{watermelon}. This indicates that the fate of a clone is hidden in its pre-perturbation state, and \CloneOT\ reveals it by predicting the response of the clone.

To test whether \CloneOT\ also forecasts these programs before treatment, we split the test clones of each scenario into thirds by their predicted program score (Fig.~\ref{fig:programs}c). The interferon program (PC10) is led by \textit{OASL}, \textit{CCL5}, \textit{BST2} and \textit{ISG15}. PC6 separates interferon-response genes from antioxidant genes such as \textit{TXN}, \textit{NQO1} and \textit{PRDX1}. For both programs, the observed score rises clearly from the bottom to the top third of the \CloneOT\ prediction. The predictions without lineage and the scores before treatment separate the clones much less. \CloneOT\ therefore forecasts which clones will induce the interferon program and which the antioxidant program, the two programs that separate non-cycling from cycling persisters.

%% file: MainText/discussion.tex
In this work, we develop \CloneOT\ as a neural inverse optimal transport method. \CloneOT\ learns the transport cost of single-cell perturbation responses from partial lineage information. Key features distinguishing it from previous methods include (1) a neural transport cost learned from clone sets, with an objective that generalizes the classical inverse OT likelihood from pairs to sets of cells, and (2) a neural entropic OT solver with closed-form conjugates and an envelope gradient, so the cost is learned without unrolling Sinkhorn iterations or map updates. Unlike previous uses of lineage tracing, \CloneOT\ uses lineage to supervise the transport cost itself. More interestingly, all cells share the learned cost, so the information of the labeled clones transfers to clones without labels. Another advantage of \CloneOT\ is its speed, since one model trains in a median of \trainminutes\ minutes on a single GPU.

Through extensive experiments, we demonstrate that \CloneOT\ recovers the true transport on two 2D datasets, while the squared Euclidean cost fails. On 14 lineage-tracing scenarios, \CloneOT\ predicts clone-level responses more accurately than five baselines, and it captures the differences between clones. In Watermelon, it is the only method that predicts clone means more accurately than the strong Observed baseline \cite{ahlmann2025deep,systema}. In the scenario of drug-tolerant persister cells, \CloneOT\ forecasts from the pre-perturbation state which clones will become cycling persisters, clearly better than without lineage. It also forecasts which clones will induce the interferon and antioxidant programs of the two fates \cite{watermelon}, although these programs are not upregulated before treatment.

As is previously highlighted, \CloneOT\ learns the transport cost from clone sets, a weak form of supervision that does not pair individual cells. A potential future development is therefore to add lineage-tracing experiments to the training data of virtual cell models \cite{bunne2024virtual,roohani2025virtual,adduri2025state}. These models currently learn perturbation responses from unpaired populations, and our results show that lineage improves the accuracy even when only 10\% of the clones are labeled. Benefited from its neural formulation, \CloneOT\ can also be extended to condition the cost on drugs, doses and time points, similar to conditional OT maps \cite{cmonge}.

In summary, \CloneOT\ represents a widely applicable framework for learning transport costs from lineage tracing. It also provides a practical training scheme for neural inverse optimal transport. Considering the fact that lineage tracing is increasingly combined with perturbation screens, spatial profiling and multimodal measurements, we believe that \CloneOT\ will be helpful wherever groups of observations are known to correspond while individual pairings remain hidden.